\documentclass[acmsmall,screen,review,nonacm]{acmart}
\renewcommand\footnotetextcopyrightpermission[1]{} 
\usepackage{import}
\usepackage{amsthm}
\usepackage{amsmath}
\usepackage{xspace}
\usepackage{subfigure}
\usepackage{makecell}

\newcommand\etal{\emph{et al.\ }}

\newtheorem{definition}{Def}
\newtheorem{corollary}{Col}

\numberwithin{definition}{section}
\numberwithin{corollary}{section}

\newcommand{\kw}[1]{{\it #1}}	

\newcommand{\kn}[1]{\textsf{\small #1}}	
\newcommand{\kd}[1]{\textbf{#1}}

\newcommand{\pt}{PTree\xspace}
\newcommand{\rt}{RTree\xspace}

\newcommand{\cirT}{$\mathcal{C}_T$} 
\newcommand{\cir}{$\mathcal{C}$}
\newcommand{\Qset}{$\mathcal{Q}$}

\newcommand{\src}{$\mathcal{C}_{Ts}$\xspace}
\newcommand{\dst}{$\mathcal{C}_{Tt}$\xspace}

\newcommand{\pat}{$\mathcal{R}$}

\newcommand{\sem}[1]{{$[\![$#1$]\!]$ }}	

\newcommand{\Rz}{\textsf{\small R}$_z$\xspace}

\usepackage[ruled,linesnumbered, vlined]{algorithm2e} 
\SetKw{Return}{return}
\SetKwData{Context}{Context}
\SetKwData{newC}{newContext}
\SetKwData{Succ}{child}
\SetKwData{Sibling}{nextSibling}
\SetKwData{find}{find}
\SetKwData{finalRule}{$R_f$}
\SetKwFunction{PM}{Match}
\SetKwFunction{Scheduler}{RuleScheduler}
\SetKwFunction{IsOpt}{isPrefix}
\SetKwFunction{Rewrite}{Rewrite}
\SetKwInOut{Input}{Input}
\SetKwInOut{Output}{Output}
\SetKwProg{Fn}{Function}{\string:}{End\ Function}
\SetKwComment{tcp}{$\vartriangleright$ }{}

\SetKwFunction{PTG}{PatternTreeGeneration}

\SetKwRepeat{Do}{do}{while}

\SetCommentSty{mycommfont}

\setcopyright{none}
\copyrightyear{2024}
\acmYear{2024}
\acmDOI{XXXXXXX.XXXXXXX}

\begin{document}
\author{Mingyu Chen}
\email{id170631@mail.ustc.edu.cn}
\affiliation{%
  \institution{University of Science and Technology of China}
  \country{China}
}

\author{Yu Zhang}
\email{yuzhang@ustc.edu.cn}
\affiliation{%
  \institution{University of Science and Technology of China}
  \country{China}
}
\thanks{Corresponding author: Yu Zhang, Email: yuzhang@ustc.edu.com}

\author{Zhaoyu Zheng}
\email{zh20011231@mail.ustc.edu.cn}
\affiliation{%
  \institution{University of Science and Technology of China}
  \country{China}
}

\author{Yongshang Li}
\email{lys316@mail.ustc.edu.cn}
\affiliation{%
  \institution{University of Science and Technology of China}
  \country{China}
}

\author{Haoning Deng}
\email{denghn@mail.ustc.edu.cn}
\affiliation{%
  \institution{University of Science and Technology of China}
  \country{China}
}


\title{Pattern Tree: Enhancing Efficiency in Quantum Circuit Optimization Based on Pattern-matching}



\begin{abstract}
Quantum circuit optimization is essential for improving the performance of quantum algorithms, particularly on Noisy Intermediate-Scale Quantum (NISQ) devices with limited qubit connectivity and high error rates.
Pattern matching has proven to be an effective technique for identifying and optimizing subcircuits by replacing them with functionally equivalent, efficient versions, including reducing circuit depth and facilitating platform portability. 
However, existing approaches face challenges in handling large-scale circuits and numerous transformation rules, often leading to redundant matches and increased compilation time.
In this study,  we propose a novel framework for quantum circuit optimization based on pattern matching to enhance its efficiency. Observing redundancy in applying existing transformation rules, our method employs a pattern tree structure to organize these rules, reducing redundant operations during the execution of the pattern-matching algorithm and improving matching efficiency. 
We design and implement a compilation framework to demonstrate the practicality of the pattern tree approach. 
Experimental results show that pattern-tree-based pattern matching can reduce execution time by an average of 20\% on a well-accepted benchmark set.
Furthermore, we analyze how to build a pattern tree to maximize the optimization of compilation time. The evaluation results demonstrate that our approach has the potential to optimize compilation time by 90\%.
\end{abstract}





\maketitle
\pagestyle{plain}

\section{Introduction}

Quantum computing has made significant progress in recent years,  driven by advancements in stable qubits, programming languages, and algorithms. However, the development of quantum computers is still limited by critical factors such as the number of qubits, their stability, and the precision with which they can be manipulated. 
Following years of substantial investment in quantum hardware research and development, several devices have been successfully developed, such as IBM's 127-qubit Eagle\cite{ibm-osprey}, China's 66-qubit Zuchongzhi\cite{wu2021zuchongzhi} and QuEra's 256-qubit neutral-atom-based Aquila\cite{wurtz2023aquila}.

In the current Noisy Intermediate-Scale Quantum (NISQ) era~\cite{preskill2018quantum}, the limited number of qubits and the lengthy execution times of quantum operations significantly impact the fidelity of results. Longer execution times exacerbate decoherence and increase error rates, undermining the effectiveness of quantum computations.  
Developing efficient algorithms is crucial for mitigating these issues. 
However, as quantum applications and algorithms scale up, a more comprehensive and flexible approach is required. Optimizing and transforming quantum programs to meet these scalability requirements is essential for enhancing computational efficiency and managing error rates.

Additionally, the field of quantum computing is witnessing the emergence of various technologies such as superconducting~\cite{ibm-osprey}, ion-trap~\cite{pino2021demonstration}, and neutral atom~\cite{wurtz2023aquila}. Each of these technologies has its unique characteristics and operational requirements. This diversity introduces additional complexity, particularly in the transplantation of quantum programs across different platforms. Ensuring compatibility and optimizing performance across various systems is a critical challenge that must be addressed. 

To optimize and transplant quantum programs, pattern matching and transformation (PMT) techniques have emerged as promising solutions~\cite{nam2018automated,xu2022quartz,iten2022exact,chen2022qcir, xu2023synthesizing, li2024quarl}. 
Early quantum circuit optimization methods based on PMT typically employed expert-designed transformation rules to process input circuits straightforwardly. This approach has been adopted by many commercial quantum compilers, such as Qiskit\cite{qiskit2024}, t|ket>\cite{sivarajah2021tket}, quilc\cite{quilc2024}, as well as many academic works\cite{nam2018automated,iten2022exact,chen2022qcir}.
Recently, quantum circuit optimization based on PMT (Pattern Matching Tree) has shifted towards search-based methods, which generally explore the search space of equivalent circuits in a heuristic manner. As a representative, Quartz\cite{xu2022quartz} employs a cost-based backtracking search algorithm to apply automatically generated and validated transformation rules in order to discover an optimized circuit. 
The aforementioned approaches involve replacing specific subcircuits that match a predetermined pattern with functionally equivalent, but more efficient or hardware-appropriate versions. This process centers on transformation rules, which comprise a \kw{source pattern} (the subcircuit to be optimized) and a \kw{target pattern} (the optimized replacement). 

\subsection{Motivation}
\label{sec:motv}
With the advancement of quantum computing, reducing the compilation time for quantum programs has become a new challenge. In the NISQ era, variational algorithms\cite{tilly2021variational} have become mainstream. The main feature of these algorithms is that computation is interleaved with compilation since the parameters used in each iteration are determined by the results of the previous iteration's computation. At the same time, many vendors offer cloud services for their quantum computers, such as Qiskit\cite{qiskit2024}. The demand for online compilation imposes constraints on the compilation time for quantum programs.

To reduce compilation time, the adoption of PMT in quantum circuit optimization has to face the following two challenges. The first one is scalability. Currently, large-scale circuits have the potential to span thousands of lines (one line of one qubit). In the meantime, with the introduction of new high-level quantum gates and the new hardware gate supported by new underlying platform technology, the number of transformation rules is also increasing. The other one is algorithm efficiency. The pattern matching problem on quantum circuits is proved to be NP-complete\cite{iten2022exact}, optimizing the matching algorithm can only reduce compilation time to a limited extent.

\begin{figure}[htb]
    \centering
    \subfigure[Motivating example: A better matching strategy]{
        \centering
        \includegraphics[width=0.5\textwidth]{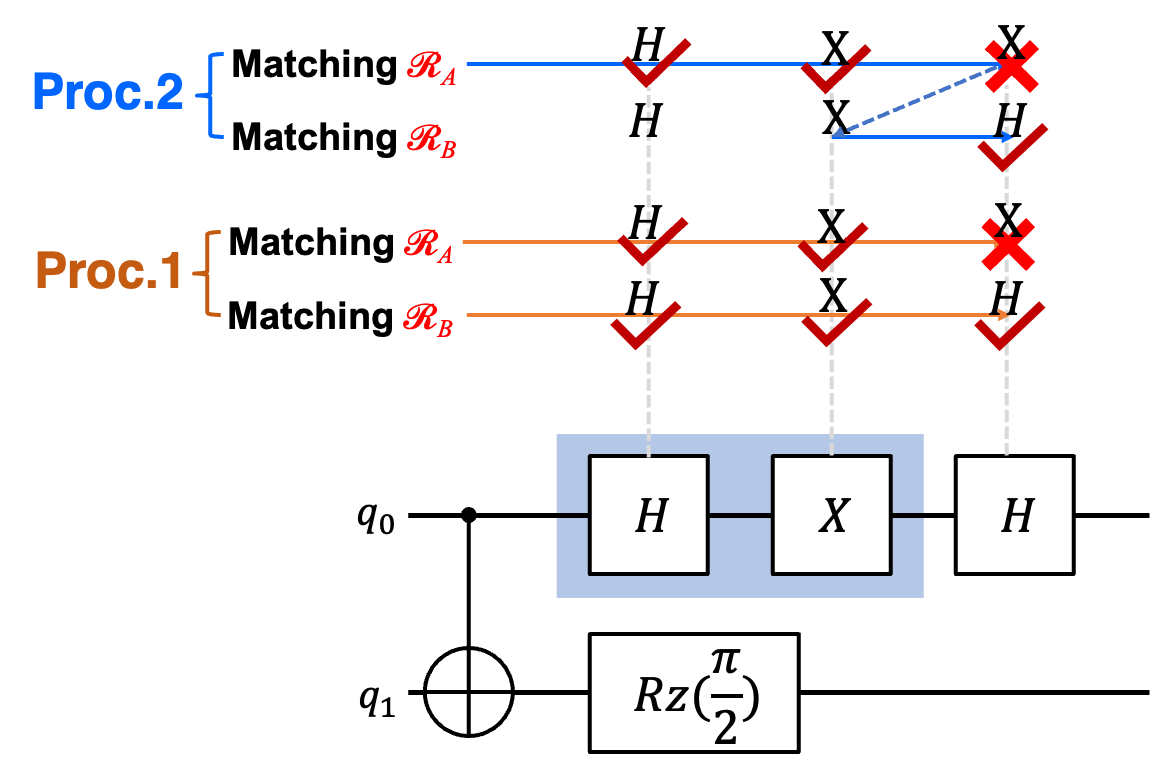}
        \label{fig:motv-ex}
    }
    \subfigure[Transformation rules A and B. Their source patterns have the same prefix subcircuit.]{
        \centering
        \includegraphics[width=0.5\textwidth]{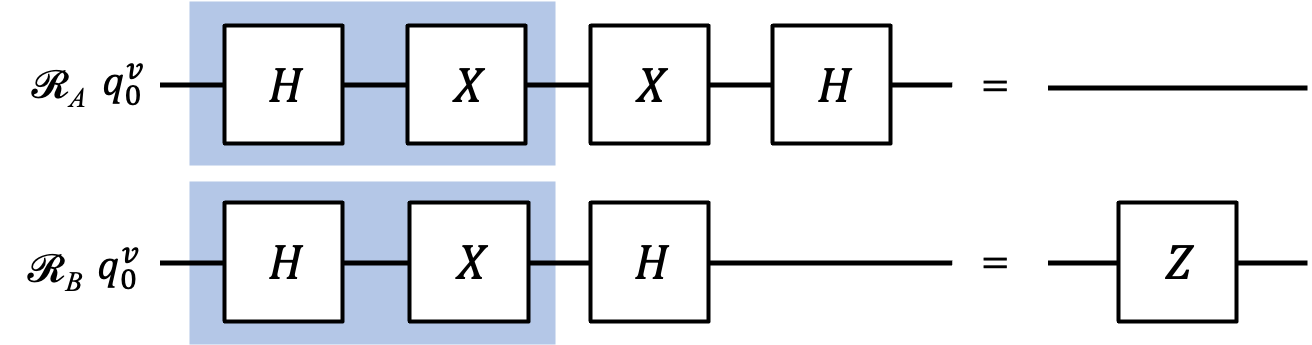}
        \label{fig:motv-a}
    }
    \subfigure[Transformation rules C and D. C's source pattern is the prefix of D's source pattern.]{
        \centering
        \includegraphics[width=0.26\textwidth]{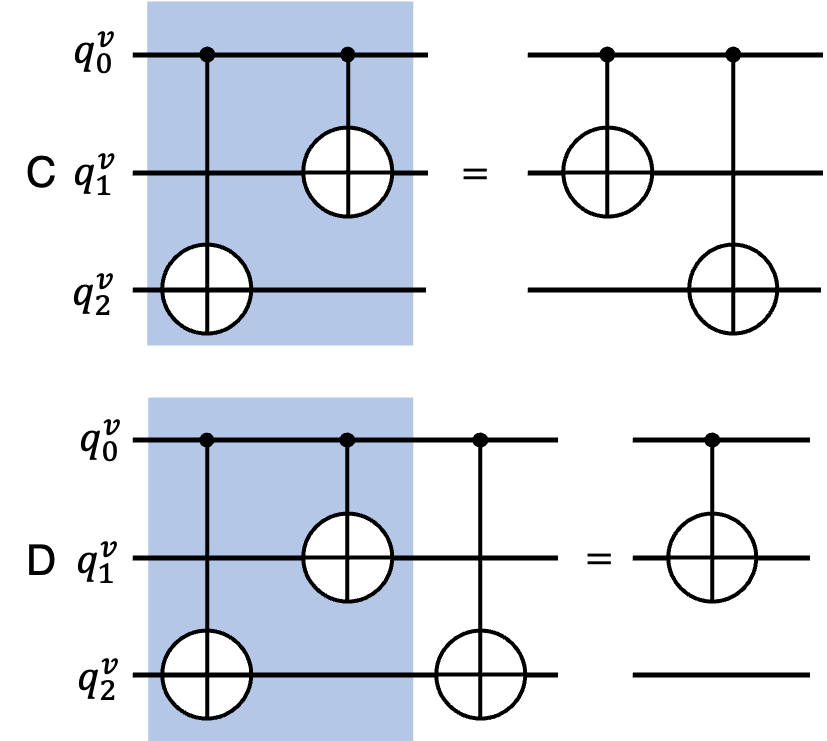}
        \label{fig:motv-b}
    }
    \caption{Motivating examples - relationships between transformation rules}
    \label{fig:motv}
\end{figure}

We use two examples to intuitively explain our idea. Fig.\ref{fig:motv-ex} shows two different matching procedures using the two rules in Fig.\ref{fig:motv-a} on a quantum circuit. In Proc.1, the compiler matches rule $\mathcal{R}_A$ firstly, and it starts matching gate by gate from the first H gate and fails to match at the third gate. Then the compiler handles rule $\mathcal{R}_B$ and succeeds. However, we notice that these two procedures matched the first two gates repeatedly. Proc.2 presents a more ideal matching: when the matching of rule $\mathcal{R}_A$ fails, instead of abandoning the matching context, the compiler can inherit the matching context and only matches the different third gate. In this way, we can reduce the cost of matching.

There is a similar problem in matching with rule $\mathcal{R}_C$ and rule $\mathcal{R}_D$. 
For an input circuit with the form the same as the source circuit of rule $\mathcal{D}$. If we apply rule $\mathcal{C}$ firstly, we need to apply an extra rule to achieve the optimization effect brought by the application of rule $\mathcal{R}_D$.

To the best of our knowledge, no work has specifically addressed this issue. Some studies have partially tackled this problem. For instance, Iten et al.\cite{iten2022exact} employed implement pattern matching by using templates, which represent a pair of equivalent circuits with a circuit identity. Their algorithm provides the maximum match for a given template, meaning that a single template can handle both rule C and rule D at the same time. However, templates cannot address the first scenario. Our work is the first to investigate the aforementioned matching redundancy and propose a solution.

\subsection{Contributions}
In this paper, we summarize the above intuitive ideas and propose \kw{pattern tree}. The pattern tree is an abstract structure used to specify the relation among transformation rules.
The pattern tree enables a more efficient organization of transformation rules, and we design a compilation framework to utilize a pattern library based on pattern tree. To verify the effectiveness of this idea, we implement a framework with the existing work Quartz\cite{xu2022quartz} to generate the transformation rules we need and QCiR\cite{chen2022qcir} to provide a pattern-matching framework prototype. Evaluation results demonstrate that the pattern tree significantly optimizes compilation time.

Our contributions are outlined as follows:
\begin{itemize}
    \item We analyze the quantum circuit transformation rules and identify the redundancy in their matching procedure. 
    \item We define a pattern tree to clarify the matching redundancy formally and design a supporting framework that applies the pattern library based on the pattern tree to optimize the quantum circuit, thereby reducing matching redundancy.
    \item We implement this framework with the existing framework Quartz and QCiR, and demonstrate its efficacy on a well-accepted benchmark set. Our evaluation shows that the pattern-tree-based method reduces compilation time by an average of 20\%. 
    \item We analyze the execution details of the pattern-tree-based pattern-matching algorithm. We have demonstrated that by adjusting the number of prefix rules, we can optimize the compilation time by up to 90\%.
\end{itemize}

The rest of this paper is organized as follows. 
Section \ref{sec:pre} clarifies the fundamentals of quantum circuit pattern matching and reviews related works.
Section \ref{sec:des} defines the pattern tree and designs a compilation framework to utilize the new pattern-tree-based transformation rules. Section \ref{sec:pt} introduces the implementation details of the pattern tree. Section \ref{sec:eval} shows the evaluation results, and the related work is discussed in Section \ref{sec:rel}.
\section{Preliminaries}
\label{sec:pre}
A quantum circuit is a common abstract model for quantum computation in which a computation is a sequence of quantum gates over a set of qubits. In quantum circuit, quantum information is transmitted along horizontal qubit wires, which denote time evolution of qubits with time propagating from left to right. The circuit depth is the length of the longest path from the input to the output, while the depth of a quantum gate in circuit is the length of the longest path from the input to the gate itself.
We formalize quantum circuit as follow:
\begin{definition}[Quantum Circuit \cir]
A quantum circuit $C = (G, $\Qset$)$ includes following parameters:
\begin{itemize}
    \item A gate sequence $G=g_0g_1...g_{n-1}$ includes all quantum gates in circuit. $n$ is the gate count, and $g=($\kn{op}$,$\kn{qlist}$)$ represents a quantum gate. \kn{op} and \kn{qlist} describes the operation type of quantum gate and the operator qubits respectively.
    \item A qubit list \Qset$ = \{q_0, q_1,...,q_{m-1}\}$.  $m$ is qubit count.
\end{itemize}
To ensure the uniqueness of the gate sequence for each quantum circuit, special constraints are added:
\begin{enumerate}
    \item For $g_i, g_j$, if $i < j$, then dep$(g_i)\le$  dep$(g_j)$, where dep$(g)$ is the depth of gate $g$;
    \item For $g_i, g_j$, if $i < j$ and dep$(g_i) = $ dep$(g_j)$, then min\_index$(g_i)$ < min\_index$(g_j)$, where min\_index$(g)$ is the minimun index of qubits in $g.$\kn{qlist}.
\end{enumerate}
\end{definition}

\begin{figure}[htbp]
    \centering
    \subfigure[Quantum circuit pattern]{
        \includegraphics[width=0.42\textwidth]{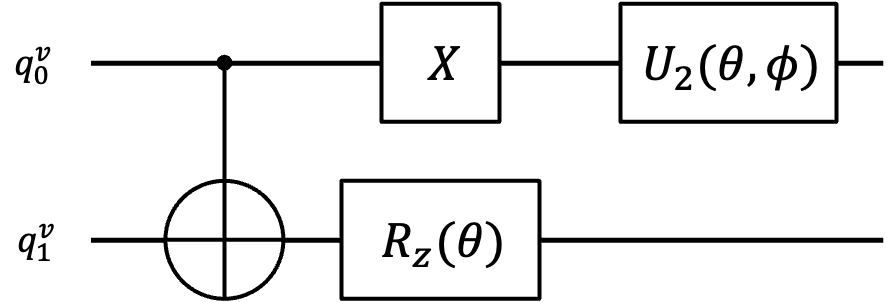}
        \label{fig:cir-ct}
    }
    \subfigure[Quantum circuit]{
        \includegraphics[width=0.42\textwidth]{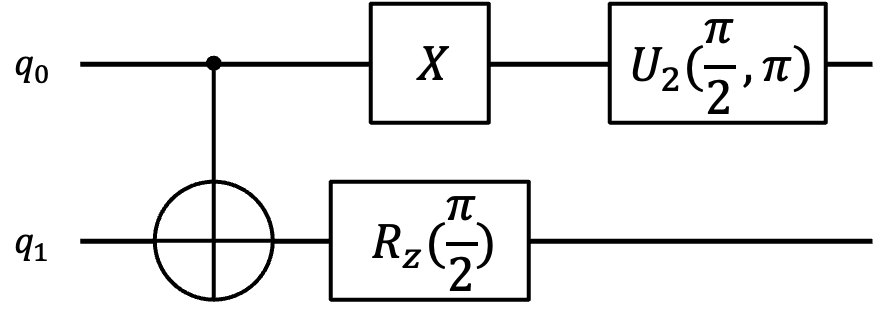} 
        \label{fig:cir-c}
    }
    \\
    \subfigure[An example of transformation rule in \cite{nam2018automated} ]{
        \includegraphics[width=0.86\textwidth]{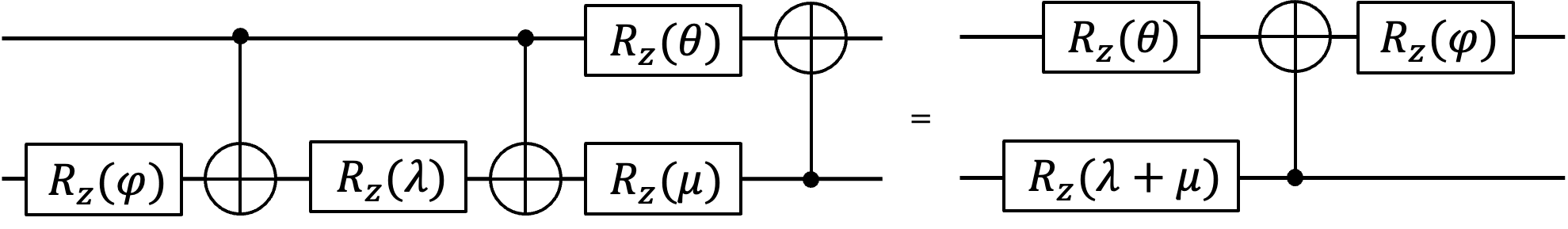}
        \label{fig:cir-rule}
    }
    \caption{Illustrating quantum circuit pattern and transformation rule}
    \label{fig:cir}
\end{figure}

\begin{example}
Fig.\ref{fig:cir-c} is an example of quantum circuit. Its gate sequence is $G = $
\kn{CNOT} q$_0$,q$_1$;
\kn{X} q$_0$;
\kn{R}$_z$($\frac{\pi}{2}$) q$_1$;
\kn{U}$_2$($\frac{\pi}{2}$,$\pi$) q$_0$, and its qubit list is $Q=\{q_0, q_1\}$.
\end{example}

It is a convenient way to represent the quantum gate in matrix format\cite{nielsen2010quantum:qcqi}. Matrix format can be used to compute how quantum gates operate on the input quantum state. For example, the matrix for the circuit shown in Fig.\ref{fig:cir-c} is (\kn{CNOT}) $\cdot$ (\kn{X} $\otimes$ \kn{R}$_z$($\frac{\pi}{2}$)) $\cdot$ (\kn{U}$_2$($\frac{\pi}{2}$,$\pi$) $\otimes$ \kn{I}).

To support pattern matching for quantum circuits, we define circuit pattern as follows.
\begin{definition}[Circuit Pattern $\mathcal{C}_T$]
A circuit pattern $\mathcal{C}_T = (G, $\Qset $_v, \vec{v}_a)$ over $m$ abstract qubits and $p$ parameters includes following parameters.
\begin{itemize}
    \item A gate sequence $G=g_0g_1...g_{n-1}$ includes all quantum gates in circuit pattern. Its difference from circuit is $g_i.$\kn{op} may include parameters.
    \item A abstract qubit list \Qset$_v = \{q_0^v, q_1^v,...,q_{m-1}^v\}$. $q_i^v(i=0,...,m-1)$ is abstract qubit, which is introduced to distinguished from the qubit in the circuit. $m$ is qubit count.
    \item A parameter vector $\vec{v}_a = \{\theta, \phi, ...\}$($|\vec{v}_a| = p$) includes all parameters in the operation type of all gates in gate sequence. 
\end{itemize}
Because the qubits in circuit pattern are abstract, circuit pattern must be \kw{instantialized} before it is used as a subcircuit, which means the abstract qubits is assigned with qubit in circuit, and each parameter is assigned with a value. 
With parameters and abstract qubits, circuit pattern can represent a set of circuits.
\end{definition}

\begin{example}
\label{exp:1} 
For a circuit pattern \cirT, we get circuit $\mathcal{C}_T\langle Q,\vec{v}\rangle$ by instantializing \cirT with qubit list $Q$ and vector $\vec{v} \in \mathbb{R}^p$. 
For example, for circuit pattern \cirT shown in Fig.\ref{fig:cir-ct}, with assigning $\{q_0^v, q_1^v\}$ to $\{q_0,q_1\}$ and assigning $(\theta,\phi)$ to $(\frac{\pi}{2},\pi)$, we instantialize \cirT with $Q=\{q_0,q_1\}$ and $\vec{v}=\{\frac{\pi}{2}, \pi\}$ to get circuit $\mathcal{C}_T\langle Q,\vec{v}\rangle$ shown in Fig.\ref{fig:cir-c}.
\end{example}

Since pattern matching on quantum circuit is based on functional equivalence between circuit patterns, similar to Quartz\cite{xu2022quartz}, we present denotation \sem{\cirT} to obtain semantics of circuit pattern \cirT. For a circuit pattern \cirT over $m$ abstract qubits and $p$ parameters, \sem{\cirT} is a $2^m \times 2^m$ unitary matrix with $p$ parameters. 
Based on semantics of circuit pattern, we can define transformation rules to describe equivalence between circuit patterns. We use \sem{\cirT$\langle\vec{v}\rangle$} to avoid different alphabets affecting the comparison between parametric matrices.

\begin{definition}[Transformation rule \pat]
A transformation rule \pat is composed of a source circuit pattern \src and a target circuit pattern \dst. The source pattern \src describes the circuit pattern to be matched, while the target pattern \dst is the circuit pattern will be used to replace the matched subcircuit within the input circuit. 
\src and \dst satisfy the following constraint
\[
\exists \beta \in \mathbb{R}, \forall \vec{v} \in \mathbb{R}^p, [\![ \mathcal{C}_{Ts}\langle\vec{v}\rangle ]\!] = e^{i\beta }[\![ \mathcal{C}_{Tt}\langle\vec{v}\rangle ]\!]
\]
$\mathcal{R}$ is the set of real numbers. That is, two circuit pattern are equivalent if, for every assignments of parameters, the semantics differ only by a phase factor.
\end{definition}


The measurement results of circuit won't be affected by the global phase, because they give the exact same distributions for any measurement in any basis. 
Besides, this relaxation is conducive to optimization. For example, the framework proposed by Maslov \etal\cite{maslov2016advantages} replace multi-control Toffoli gates with simpler relative-phase implementation while preserving the functional correctness for optimization. 

Furthermore, the semantics of quantum circuit is a $2^m \times 2^m$ unitary complex matrix. Now we can compare the semantics of instantialized circuit pattern with circuit. 
\begin{definition}[Match]
A circuit \cir \kd{matches} a circuit pattern \cirT, if 
\[
\exists Q \in \mathbb{Q}^m, \vec{v}\in \mathbb{R}^p, s.t. [\![\mathcal{C}_T\langle Q, \vec{v}\rangle]\!] = [\![C]\!]
\]
where $\mathbb{Q}$ represents qubit in \cir$.$\Qset. That is, \cir matches \cirT if there is an assignment of abstract qubit list and parameter vector of \cirT that makes the semantics of \cir and \cirT equivalent. In the rest of the paper, $Q$ and $\vec{v}$ are called \kd{matching context}.
\end{definition}

\section{Design}
\label{sec:des}

\begin{figure*}[htb]
    \centering
    \subfigure[PTree]{
        \centering
        \includegraphics[width=0.35\textwidth]{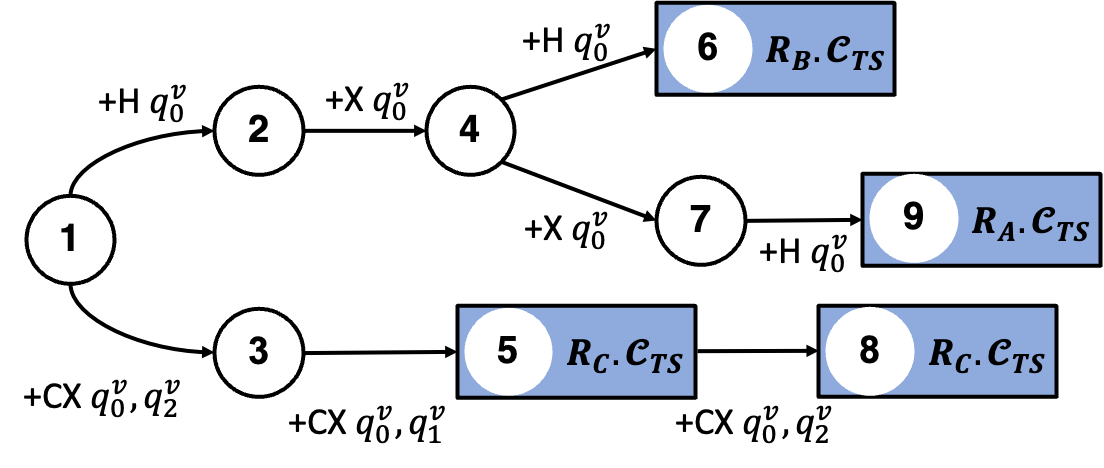}
        \label{fig:tree-pt}
    }
    \subfigure[RTree]{
        \centering
        \includegraphics[width=0.28\textwidth]{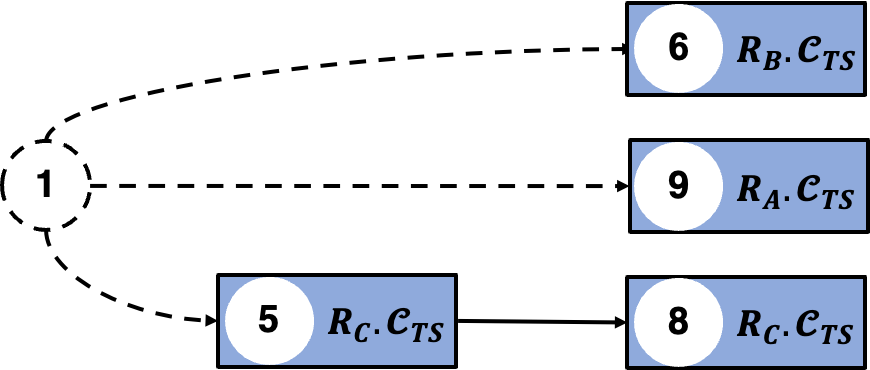}
        \label{fig:tree-rt}
    }
    \subfigure[RTree with special nodes]{
        \centering
        \includegraphics[width=0.3\textwidth]{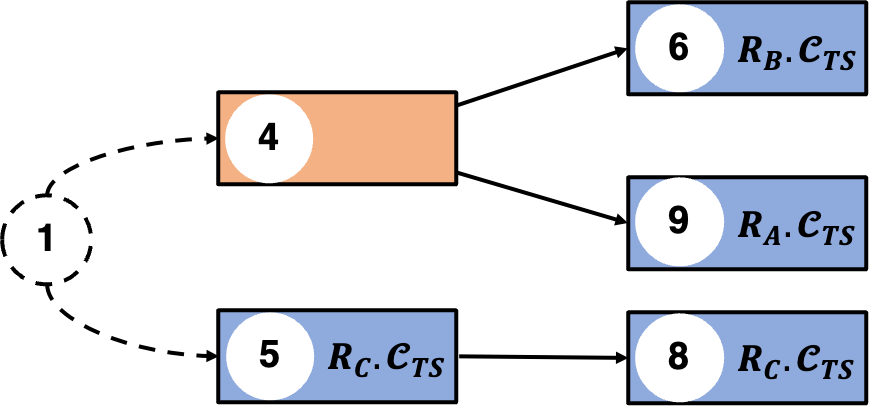}
        \label{fig:tree-sprt}
    }
    \caption{The variants of pattern tree}
    \label{fig:tree}
\end{figure*}

To address the discussed problems, we define a pattern tree, which is an abstract tree data structure for describing quantum circuit transformation rules. 
In this section,  we first define two variants of pattern tree \pt and \rt. 
Then, we introduce the tailored compilation framework for processing pattern library based on pattern tree.

The following terminologies will be used from now on:
(a) \kw{source} is the circuit pattern to be matched;
(b) \kw{input} is the input circuit within which the pattern-matching algorithm will detect optimization opportunities;
(c) \kw{target} is the circuit pattern that will be used to replace the matched subcircuit in the input.

\subsection{\pt and \rt}
\label{subsec:sr}
The definition of \rt is based on the precedence between the circuit patterns. 

\begin{definition}[Circuit precedence]
For circuits \cir$_a$ and \cir$_b$, the gate sequence \cir$_a.G$ $=g_{0}^a g_{1}^a ...g_{k-1}^a$ precedes \cir$_b.G$ $=g_{0}^b g_{1}^b...g_{l-1}^b$, written as \cir$_a.G\prec$\cir$_b.G$, if $|$\cir$_a.G|=k<l=|$\cir$_b.G|$ and \cir$_a.G$ is equivalent to \cir$_b.G$'s subsequence $G'=g_{0}^b...g_{k-1}^b$.
\end{definition}

\begin{corollary}
Similarly, circuit pattern precedence is defined: for circuit patterns $\mathcal{C}_{Ta}$ and $\mathcal{C}_{Tb}$, the gate sequence $\mathcal{C}_{Ta}.G$ $=g_{0}^a g_{1}^a...g_{k-1}^a$ precedes $\mathcal{C}_{Tb}.G$ $=g_{0}^b g_{1}^b ...g_{l-1}^b$, written as $\mathcal{C}_{Ta}.G\prec$$\mathcal{C}_{Tb}.G$, if $|\mathcal{C}_{Ta}.G|=k<l=|\mathcal{C}_{Tb}.G|$ and $\mathcal{C}_{Ta}.G$ is equivalent to $\mathcal{C}_{Tb}.G$'s subsequence $G'=g_{0}^b...g_{k-1}^b$.
\end{corollary}

For simplicity, we refer to both types of precedence as circuit precedence.

As introduced in Sec.\ref{sec:pre}, each transformation rule is composed of source and target. Based on circuit precedence, the sources of transformation rules can form a tree, which is \kw{\pt}.

\begin{definition}[PTree $\mathcal{T}_P=(V,E)$]
\label{def:pt}
In \pt  $\mathcal{T}_P=(V,E)$, each node $v\in V$ represents a circuit pattern. Each directed edge $e =(u,v,g)\in E$ means that the circuit pattern represented by node $v$ is derived from adding gate $g$ to the circuit pattern represented by node $u$.
\end{definition}

\begin{example}
We can use the sources of transformation rules shown in Fig.\ref{fig:motv} to construct a \pt, as illustrated in Fig.\ref{fig:tree-pt}. 
The node 1, which is the root of this \pt, represents a null circuit pattern.
Node 9,6,5,8 are respectively the sources of the rules A,B,C,D.
We can simply identify the longest common prefix circuit of different rules in the \pt.
Node 4 represents the longest common prefix circuit of sources of rule A and rule B, and node 5 is the source of rule C and also the longest common prefix circuit for the sources of rule C and rule D.
\end{example}

For simplicity, we will refer to the prefix circuit of the sources of the transformation rules as the prefix circuit of the transformation rules.

In Def.\ref{def:pt}, notice that each node represents circuit pattern instead of transformation rules. The reason is that only the sources of the transformation rules are used during the pattern matching process. We don't need to care about what target is during matching. 
Moreover, not all circuit patterns can serve as sources for transformation rules. This can accommodate scenarios where one source corresponds to multiple targets.
We say that a node whose corresponding circuit pattern can be used as source is called \kw{valid}. 

Through an appropriate implementation, multiple transformation rules can be organized into a \pt. The pattern matching framework can handle multiple rules simultaneously using the \pt structure.
If the match for current rule succeeds, the algorithm can check whether it has child nodes for advancing. 
If the match for current node fails, the algorithm can backtrack for other choices.
However, \pt needs a large amount of nodes to describe the relationships between transformation rules. To reduce the consumption of storage space, we propose \rt.

\begin{definition}[\rt $\mathcal{T}_R=(V,E)$]
In \rt $\mathcal{T}_R=(V,E)$, each node $v\in V$ represents a circuit pattern, each directed edge $e = (u,v)$ indicates that the circuit pattern of node $u$ precedes the circuit pattern of node $v$.
\end{definition}

\begin{example}
The \rt shown in the Fig.\ref{fig:tree-rt} is also built form the transformation rules in Fig.\ref{fig:cir-rule}. For ease of reference, we retain the node numbers from Fig.\ref{fig:tree-pt}.
The dashed-line node 1 is used to specify the relationship among rules. \rt does not contain node 1.
Similar to \pt, \rt can also demonstrate the relationships between different rules.
Their differences are as follows:
\begin{itemize}
    \item All nodes in \rt are valid. Therefore, in addition to storing transformation rules, A only needs to store the relationships(edges) between rules, while B also needs to store a large number of additional nodes. 
    \item  \rt finds it difficult to utilize common prefix of rules to reduce the redundancy in matching process. Without invalid nodes, \rt can hardly even be considered a tree. 
\end{itemize}
\end{example}

To achieve both the compilation time optimization advantages of \pt and the space-saving advantages of \rt, we choose to make a compromise -- select some nodes from \pt and add them to \rt, as shown in Fig.\ref{fig:tree-sprt}. By adding the node 4 in \pt into \rt, we can connect node 6(rule B) and node 9(rule A). It is reflected in the matching process as the matching of rule A and rule B can reuse part of the other's matching context. 

In the following content, we collectively refer to \pt and \rt as \kw{pattern tree} when differentiation between them is unnecessary.

\subsection{Pattern-matching algorithm}
\begin{figure*}[htb]
\subfigure[]{
    \includegraphics[width=0.8\textwidth]{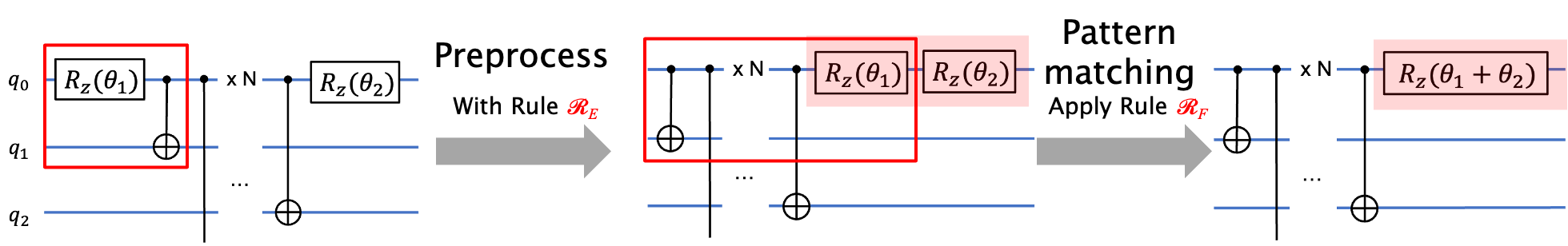}
    \label{fig:rz-flow}
}
\subfigure[Commutation rules for R$_z$ gates]{
    \includegraphics[width=0.4\textwidth]{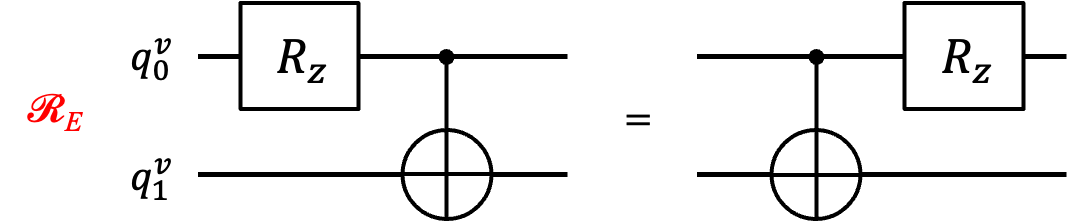}
    \label{fig:rz-com}
}
\subfigure[Merging rules for R$_z$ gates]{
    \includegraphics[width=0.4\textwidth]{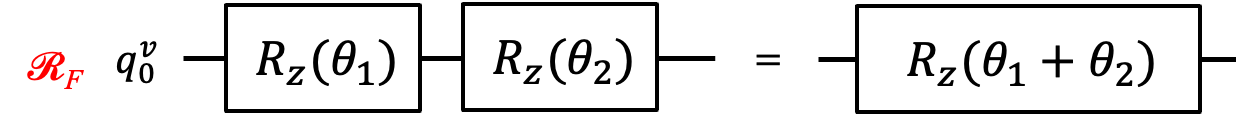}
    \label{fig:rz-merge}
}
\caption{Preprocess}
\label{fig:preprocess}
\end{figure*}
\begin{figure*}[htbp]
    \centering
    \includegraphics[width=0.8\textwidth]{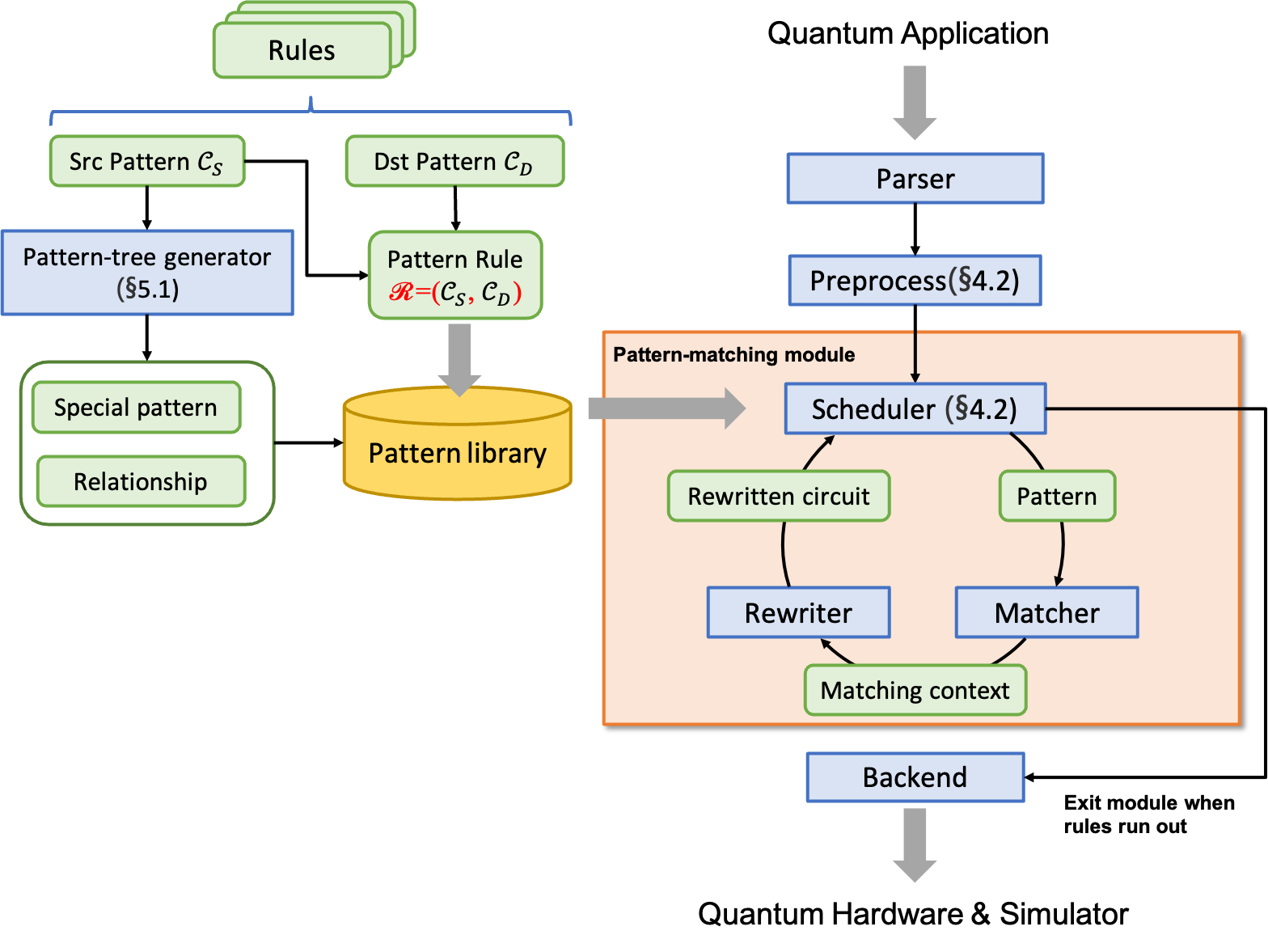}
    \caption{The compilation flow in our pattern-matching framework}
    \label{fig:frame}
\end{figure*}

In this subsection, we design a compilation framework to employ the pattern library based on pattern tree as shown in Fig.\ref{fig:frame}. The whole framework relies on the combined efforts of the pattern library and pattern scheduler.

Before the execution of pattern scheduler, pattern library need to be filled with the transformation rules. The transformation rules can be obtained from the existing work. What we need to do is to convert them into our format. Based on this, we utilize the sources of these rules to generate pattern tree structure, which includes the aforementioned special circuit patterns and the relationships among circuit patterns. Implementation details will be discussed in Sec.\ref{sec:pt}.

The other part of our framework resembles a normal compilation flow. We assume that the input to our framework is a quantum program from a quantum application, and the output will be sent to the underlying quantum hardware.
First, the frontend converts input quantum program into quantum circuit. Afterwards, the circuit is transferred to preprocess routine, then the preprocessed circuit is passed to scheduler.
The scheduler is designed to perform the following operations iteratively:
1) fetch transformation rules from pattern library; 
2) match according to the rule until find a matched subcircuit; 
3) pass matching context to the rewriter for replacing the matched subcircuit. 
Finally, when the rules in pattern library run out, the rewritten circuit will be passed to backend and transformed to a proper format for hardware.


\subsubsection{Preprocess}

Before adding preprocess routine, we observed that our compilation framework struggled to achieve the same level of optimization as existing methods. Direct matching usually fail to produce satisfactory results. 
For example, for circuits with the form shown in the leftmost side of Fig.\ref{fig:rz-flow}, the $R_z$ gates on two sides can be put together according to the commutation rule $\mathcal{R}_E$ shown in Fig.\ref{fig:rz-com}, and be merged into one $R_z$ gate according to the merging rule $\mathcal{R}_F$ shown in Fig.\ref{fig:rz-merge}.

However, if we use pattern matching to perform the above operations, we need to apply rule $\mathcal{R}_E$ $N$ times to get the two $R_z$ adjacent so that we can apply rule $\mathcal{R}_F$, or we need a rule of size N+2 which includes two $R_z$ gates and all the CNOT gates between them to directly merge $R_z$ gates in one step. 
The former requires the compiler to apply rule $\mathcal{R}_E$ multiple times, which doesn't change the cost of circuit. But if we allow the application of commutation rules, it will result in an increase in the number of transformation rules and an expansion of the search space, consequently leading to longer execution time, which is contrary to our initial intention.
The latter requires the pattern library to include the transformation rules all possible $N$, which is impossible, let alone the size of pattern library grows exponentially with $N$.
Therefore, direct application of pattern matching without any guidance is inefficient.

Therefore, we design a preprocess routine to put the $R_z$ gates which are far from each other together. Instead of aimless matching, the routine will categorize $R_z$ gates into different groups according to commutation rules. For example, since there are only target qubits of \kn{CX} gate between \Rz gates, they can be categorized into one group based on $\mathcal{R}_E$. After determining all groups, the rotation gates in the same group will be moved to adjacent positions. The converted circuit is shown as the second circuit of Fig.\ref{fig:preprocess}. The final merging step will be finished by pattern matching routine, then we will get the rightmost circuit shown in the Fig.\ref{fig:preprocess}.

\subsubsection{Rule Scheduler}
To utilize pattern-tree-based pattern library, a tailored algorithm is required.
The Alg.\ref{alg:sche} lists our rule scheduler algorithm. The input $\mathcal{G}$ is the DAG representation of the quantum program to be optimized, and $r$ is the fetched transformation rule used in this round of optimization. In Alg.\ref{alg:sche}, the input transformation rule is treated as a tree structure. Its \kn{child} function return its first child node(rule), and \kn{nextSibling} return its sibling node(rule).

\IncMargin{1em}
\begin{algorithm}
\Input{Input quantum circuit $\mathcal{C}$, a given transformation rule $r$}
\Output{Rewritten circuit $\mathcal{C}'$}
\Fn{\Scheduler{$\mathcal{C},r$}}{
    \find $\leftarrow$ \kn{False}\tcp*[r]{The boolean varibale used to specify if matching succeeds}
    \Do{\kn{True}}{
        \Context$\leftarrow$ \PM{$\mathcal{C}$, $r$}\;
        \uIf{\Context}{
            $R_f$ $\leftarrow$ $r$\;
            \find $\leftarrow$ True\;
            \uIf{$r$.\Succ}{
                $r\leftarrow$ $r$.\Succ\;
            }
            \Else{
                break\;
            }
        }
        \ElseIf{$r$.\Sibling}{
            $r\leftarrow$ $r$.\Sibling
        }
        \Else{
            break\;
        }
    }
    \If{find $\mathrm{and}$ \IsOpt{$R_f$}}{
        $\mathcal{C}$' $\leftarrow$ \Rewrite{$\mathcal{C}$,$r$,\Context}\;
        \Return{$\mathcal{C}$'}\;
    }
}
\caption{Rule scheduler for pattern matching algorithm}
\label{alg:sche}
\end{algorithm}
\DecMargin{1em}

The rule scheduler handling one transformation rule can be divided into two phases: matching and rewriting. In this work we only introduce the processing of pattern-tree-based transformation rule in matching procedure. The details of matching and rewriting are not taken into our scope.
Algorithm \ref{alg:sche} illustrates our depth-first processing procedure for pattern-tree-based rule in a non-recursive style.
In each iteration of algorithm, at first, \kn{Match} function is called to get the matching context of rule $r$ on circuit $C$(line 4). If the matching context is returned(line 5), the compiler will update the rule $\mathcal{R}_f$ finally used to apply and the boolean variable \kn{find} used to specify if the matching is successful(line 6-7). Then we check if current rule has any child. If so, we assign $r$'s child to $r$ and start a new iteration(line 8-9). Otherwise, we jump out of the loop(line 10-11).
If the \kn{Match} function fail to find matched subcircuit for rule $r$, the compiler checks if it has any sibling rules(line 12-13). If not, we also jump out of the loop(line 14-15). 

Notice that we use a function \kn{isPrefix} to check the feasibility of $R_f$. This is related to the measures for adding special nodes into pattern library, which will be discussed in detail in the next section. And after rewriting, the rewritten circuit is returned for next round of optimization.

\section{Implementation of pattern tree}
\label{sec:pt}

In this section, we discuss the implementation details of pattern tree.

As we mentioned in the Sec.\ref{subsec:sr}, the basic idea of pattern tree is to reduce the overhead of matching by utilizing common prefix circuit patterns. However, in most cases, these common prefix circuits cannot serve as the sources for any transformation rule. As a result, these patterns are difficult to include in a pattern library aimed at optimization. 

To address this problem, we choose to pack each prefix circuit pattern into a special type of rule. This rule uses the prefix pattern as its source and an empty circuit as its target, but includes an additional attribute to ensure it is used solely for matching and not for rewriting. Therefore, as shown in Algorithm \ref{alg:sche}, we add a conditional check before rewriting. In the following content, we call this special rule \kw{prefix rule}.

\begin{definition}[Prefix rule $\mathcal{R}^p$]
We say rule $\mathcal{R}^p$ is prefix rule if 
\begin{itemize}
    \item There exist a set of rule $\mathbb{R} = \{\mathcal{R}_i | i=0,...,n\}$ contain at least one rule. For each $\mathcal{R}_i$ in $\mathbb{R}$, $\mathcal{R}^p.$\src precedes $\mathcal{R}_i.$\src. 
    \item $\mathcal{R}^p$ does not have a feasible target, which means it won't be applied.
\end{itemize} 
\end{definition}


\begin{figure}[htb]
    \centering
    \includegraphics[width=0.8\textwidth]{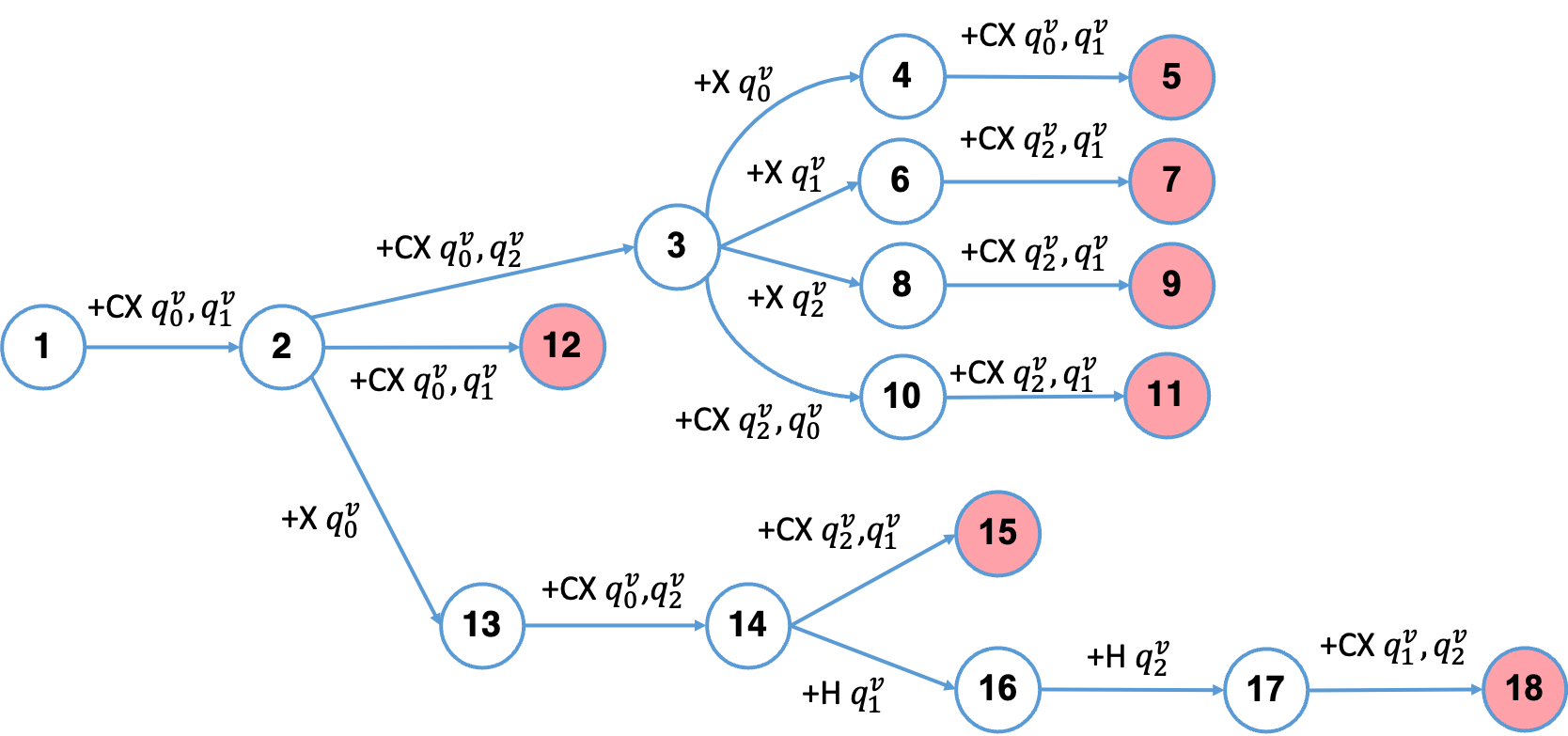}
    \caption{An example PTree used to discuss how to add special rules.}
    \label{fig:pt}
\end{figure}

The introduction of prefix rule raises two questions:
\begin{itemize}
    \item RQ1: Which prefix rules should be added to the pattern library?
    \item RQ2: What is the appropriate number of prefix rules to add to the pattern library?
\end{itemize}
We only discuss RQ1 in this section. RQ2 will be discussed in the Sec.\ref{sec:eval}.

\subsection{RQ1: Choice of prefix rule}
We use the \pt shown in the Fig.\ref{fig:pt} as an example to analyze RQ1. 
There are 7 rules and 11 candidate prefix rules in total in this \pt. We analyze several conditions to lay the groundwork for our answer to RQ1.
\begin{enumerate}
    \item \kd{Empty circuit.} Although empty circuit can be used as circuit pattern, each match start from an empty circuit. There is no need to specifically designate the empty circuit as prefix rule.
    \item \kd{The comparison of coverage.} Consider choosing node 2 vs. choosing node 3 as the prefix rule.
    The rules with node 2 as a prefix are rules 5, 7, 9, 11, and 12, while the rules with node 3 as a prefix are only rules 5, 7, 9, and 11. If we fail to match the circuit pattern of node 2, we can exclude more rules and thus saving more time. Therefore, node 2 is a better choice, and we are predisposed to choose the nodes that serve as prefixes for more rules.  
    \item \kd{Nodes with one child.} Comparison of the two cases: choosing node 13 vs. choosing node 14 as the prefix rule. The number of rules with either of them as a prefix is 2. Their differences are as follows:
    \begin{itemize}
        \item The circuit pattern represented by node 13 is shorter than that of node 14. This means that when a match fails, adding node 13 can lead to a quicker detection.
        \item However, node 14 is closer to the two rules, allowing it to retain more matching context.
    \end{itemize}
\end{enumerate}

\IncMargin{1em}
\begin{algorithm}
\caption{Prefix rule generation algorithm. $\mathcal{T}$ is a Trie used to store transformation rule. Each node of $\mathcal{T}$ represents a circuit pattern and each edge represents a symbolic quantum gate.}
\label{alg:ptt}
\Input{Transformation rule set $\mathbb{R}$, where each rule $\mathcal{R}_i(i=1,...,N_R)$ comprises source $s$ and target $t$, the number of prefix rules $N$}
\Output{Prefix rule set $\mathbb{R}^p$}
\Fn{\PTG{$\mathbb{R},N$}}{
    Initialize $\mathcal{T}$ 
    $=$ \kn{new} \kn{TrieNode}\tcp*[r]{$\mathcal{T}$ points to the root node of trie}
    \For{$i=1,\cdots,N_R$}{
        Initialize $p$ $=$ $\mathcal{T}$\;
        \For{each quantum gate $g$ in $\mathbb{R}[i].$\src}{
            \uIf{$p$ has no succeeding node at $g$}{
                $p[g]$ $=$ \kn{new} \kn{TrieNode}\;
            }
            $p$ $=$ $p[g]$\;
        }
    }
    Create a priority queue $\mathbb{Q}$ $=$ $\varnothing$ of Trie node $v$ with \kn{cover}$(v)$ decreasing\;
    \For{node $v$ on the $\mathcal{T}$}{
        Get the circuit pattern $\mathcal{C}$ represented by $v$\;
        \uIf{$|~v$.child$~|$ > 1 $\mathrm{and}$ $\mathcal{C}.size > 0$ $\mathrm{and}$ $\mathcal{C}$ is connected}{
            Add v into $\mathbb{Q}$\;
        }
    }
    Pack the circuit pattern represented by top $N$ node in $\mathbb{Q}$ into prefix rule $\mathcal{R}^p$\;
    \Return{$\mathbb{R}^p$ $=$ $\{\mathcal{R}^p\}$};
}
\end{algorithm}
\DecMargin{1em}

We conclude an algorithm(Alg.\ref{alg:ptt}) to generate prefix rules from the input transformation rules.
The algorithm takes the original transformation rule set $\mathbb{R}$ and the required number of prefix rule as input, and output the prefix rule set.
Firstly, it create a Trie $\mathcal{T}$ for collecting source circuit patterns of all rules(line 2), and insert all rules in $\mathbb{R}$ into $\mathcal{T}$(line 3-7).
The edges of $\mathcal{T}$ is labelled with the symbolic quantum gate. 
Therefore, in the algorithm, we access the next node through the quantum gates in the source of each rule(line 5-7).

\begin{figure*}[htbp]
    \centering
    \includegraphics[width=0.4\textwidth]{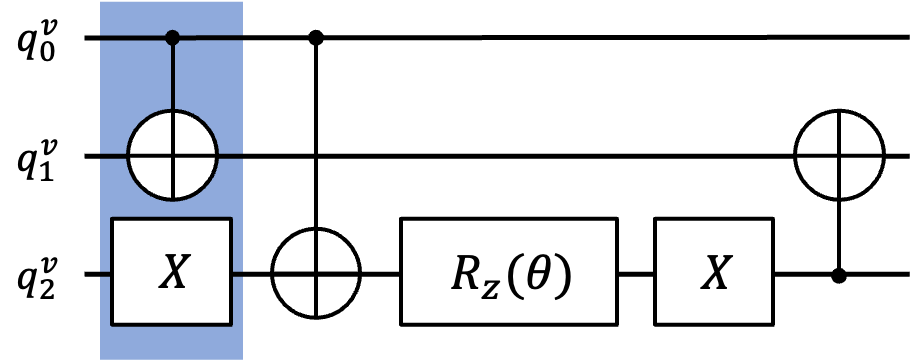}
    \caption{The unconnected subcircuit example}
    \label{fig:connected}
\end{figure*}

To answer RQ1, we define a function \kn{cover}.
\begin{definition}[cover]
We use $\mathrm{cover}$($\mathcal{C}_T$) to represent the total number of rules $\mathcal{R}$ that have $\mathcal{C}_T$ as a prefix. Prefix rule is not included.
\end{definition}
In Alg.\ref{alg:ptt}, we construct a priority queue $\mathbb{Q}$ with \kn{cover}$(r)$ decreasing as our heuristic method(line 8). We traverse $\mathcal{T}$ and insert the nodes that meet the conditions into $\mathbb{Q}$(line 10-12). These conditions are used to exclude nodes that are not suitable as prefix rule. 
\begin{enumerate}
    \item The number of children of $v$ should be greater than 1. 
    The nodes without children are the nodes represents transformation rules. They cannot be used as prefix rule.
    For the nodes with only one child, we tend to choose its child to add to the queue for alignment, since as discussed in the previous comparison, there is no significant difference between the two and exclude the one with obvious feature is easier.
    \item Using empty circuit as prefix rule is trivial.
    \item The graph representation of $C$ should be connected. The subcircuit in blue box shown in Fig.\ref{fig:connected} is an unconnected example. If we want to match this subcircuit, the loss of structure information will result in many useless matching. This is because they are connected to the same gate in original circuit, but the subcircuit does not reflect this. Therefore, we will exclude it in our prefix rule generation.
\end{enumerate}

In the end, we select the top $N$ values from $\mathbb{Q}$ as the returns(line 14). 

\section{Evaluation}
\label{sec:eval}
In this section, we first compare the performance of our framework with existing quantum circuit optimizer and demonstrate the correctness of our framework. Next we evaluate how our framework optimizes compilation time with the help of pattern tree. 

\subsection{Experiment environment}
\paragraph{Benchmarks}
We use a set of 26 benchmark circuits, which is widely adopted in existing studies\cite{nam2018automated, xu2022quartz, xu2023synthesizing, li2024quarl}. It includes Galois field multiplier circuits and arithmetic circuits and so on.

\paragraph{Metric} For optimization effect on quantum circuit, we use total gate number as the metric for optimization and transformation rules.
For optimization effect on compilation, we use the execution time of pattern matching algorithm as the main metric.

\paragraph{Transformation rules}
We use Quartz\cite{xu2022quartz} framework to generate the transformation rules we need in the experiments. 
We set maximum number of qubits and quantum gates to 3 and 6 respectively for gate set \{\kn{H}, \kn{X}, \kn{R}$_z$($\lambda$), \kn{CNOT}\} in Quartz to generate the transformation rule. The gate set \{\kn{H}, \kn{X}, \kn{R}$_z$($\lambda$), \kn{CNOT}\} is called Nam gate set in previous work. 
There are 1669 transformation rules generated originally in total. 
Notice that the number is inconsistent with the number given by Quartz. This is because we exclude the cost-preserving and cost-increasing rules during the generation procedure. Through algorithm \ref{alg:ptt}, we get 845 prefix rule candidates in total.

\paragraph{Verifier}
We use the matrix representation of quantum circuit to verify the equivalence between the circuit before optimization and after optimization.
To compute the matrix representation, we use the \kn{unitary\_simulator} provided by Qiskit\cite{qiskit2024} to get the unitary of the circuits. 
Since it only allows the circuit with less than 16 qubits, we only examine the equivalence of a subset of the benchmarks.

\paragraph{Platform}
Our program are executed on 12th Gen Intel(R) Core(TM) i9-12900K. We choose the framework QCiR\cite{chen2022qcir} to perform our experiments for two reasons: 1) It provides a relatively custom description format for users to add new transformation rule, which enables our pattern tree design;
2) It is a rule-by-rule pattern matching framework, which helps us observe the optimization effect of the pattern tree on compilation time.

Our pattern matching routine traverses on the input circuit for opportunities to greedily apply transformation rules. To avoid missing optimization opportunities due to the order of application, we performing multiple passes.

\begin{table}[ht]\small
    \centering
    \begin{tabular}{|c|c|c||ccc||cc|}
        \hline
        Circuit & Orig. & q & \makecell{Qiskit\cite{qiskit2024}} & \makecell{Quartz\cite{xu2022quartz}} & Nam\cite{nam2018automated} & Ours & Verified \\
        \hline
        adder\_8                       & 900 & 24  & 1030  & 724 & 606 & 684 & \\
        barenco\_tof\_3                & 58  & 5   & 54   & 38  & 40  & 46  & True \\
        barenco\_tof\_4                & 114 & 7   & 106  & 68  & 72  & 90  & True \\
        barenco\_tof\_5                & 170 & 9   & 158  & 98  & 104 & 134 & True \\
        barenco\_tof\_10               & 450 & 19  & 418  & 262 & 264 & 354 &  \\
        csla\_mux\_3                   & 170 & 15  & 158  & 154 & 155 & 168 & True \\
        csum\_mux\_9                   & 420 & 30  & 420  & 272 & 266 & 420 & \\
        mod5\_4                        & 63  & 5   & 61   & 26  & 51  & 56  & True \\
        mod\_mult\_55                  & 119 & 9   & 115  & 93  & 91  & 107    & True \\
        mod\_red\_21                   & 278 & 11  & 256  & 202 & 180 & 222   & True \\
        qcla\_adder\_10                & 521 & 36  & 557  & 422 & 399 & 490 & \\
        qcla\_com\_7                   & 443 & 24  & 428  & 292 & 284 & 359   &  \\
        qcla\_mod\_7                   & 884 & 26  & 886  & 719 & - & 814 &  \\
        rc\_adder\_6                   & 200 & 14  & 239  & 154 & 140 & 186   & True \\
        tof\_3                         & 45  & 5   & 43   & 35  & 35  & 35    & True\\
        tof\_4                         & 75  & 7   & 71   & 55  & 55  & 55    & True \\
        tof\_5                         & 105 & 9   & 99   & 75  & 75  & 75    & True \\
        tof\_10                        & 255 & 19  & 239  & 175 & 175 & 175   &  \\
        vbe\_adder\_3                  & 150 & 10  & 169  & 85  & 89  & 126   & True \\
        gf2\textasciicircum4\_mult     & 225 & 12  & 212  & 177 & 287 & 212    & True \\
        gf2\textasciicircum5\_mult     & 347 & 15  & 327  & 277 & 296 & 332    & True \\
        gf2\textasciicircum6\_mult     & 495 & 18  & 464  & 391 & 403 & 469    & \\
        gf2\textasciicircum7\_mult     & 669 & 21  & 627  & 531 & 555 & 634    & \\
        gf2\textasciicircum8\_mult     & 883 & 24  & 819  & 703 & 712 & 833    & \\
        gf2\textasciicircum9\_mult     & 1095 & 27 & 1023  & 873 & 891 & 1041   & \\
        gf2\textasciicircum10\_mult    & 1347 & 30 & 1248 & 1060 & 1070 & 1266 & \\
        \hline
        \multicolumn{3}{|c||}{\kd{Mean Opt. Rate}} & 2.27\% & 28.7\% & 25.10\% & 14.65\%  &\\
        \hline
    \end{tabular}
    \caption{The reduction effect taken by pattern tree on automatically generated pattern library}
    \label{tbl:eff}
\end{table}

\subsection{Optimization effect}

Table.\ref{tbl:eff} compare our framework to Qiskit\cite{qiskit2024}, Quartz\cite{xu2022quartz}, Nam\cite{nam2018automated}.
We collect the data of Quartz and Nam from their work, and collect Qiksit's data by execute the transpiler of Qiskit with setting optimization level to 3.
Although our optimization framework has optimization effect, there remains a significant gap compared to the optimization effects achieved by works such as those of Quartz and Nam. Due to the lack of consideration for non-cost-reducing rules, the optimization upper bound of our method is lower than that of search-based approaches\cite{xu2022quartz,xu2023synthesizing,li2024quarl}. In contrast to Nam's work, our framework does not fully implement all of its preprocessing steps. Since the Toffoli benchmarks relies on moving \kn{R}$_z$ gates for further optimization, our framework achieves similar performance. However, for programs that require moving H gates, our performance is not ideal.

\begin{figure}[t]
    \subfigure[The optimization rate and the compilation time]{
        \centering
        \includegraphics[width=0.85\textwidth]{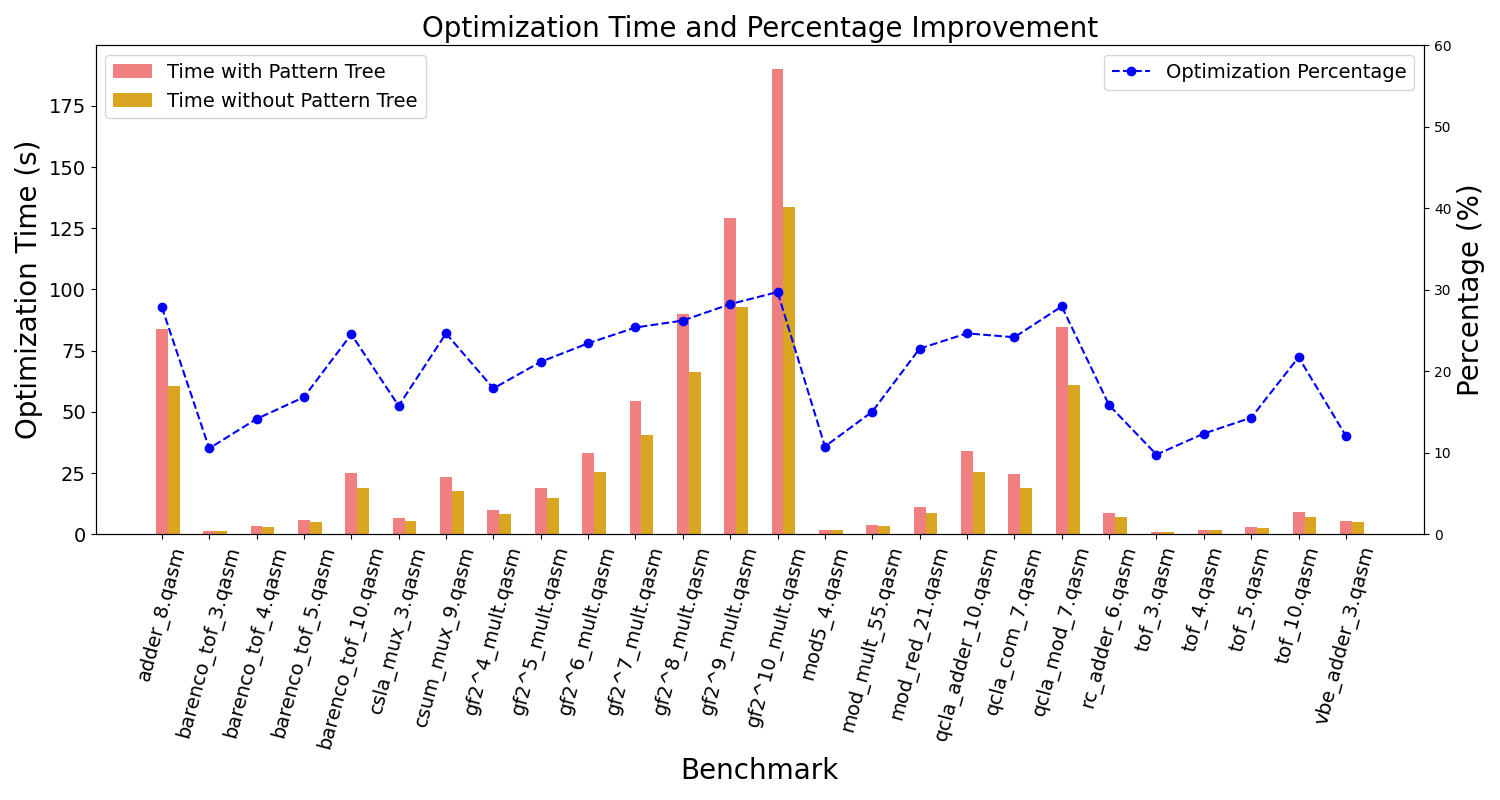}
        \label{fig:time-opt}
    }
    \subfigure[The optimization rate and the gate count]{
        \centering
        \includegraphics[width=0.85\textwidth]{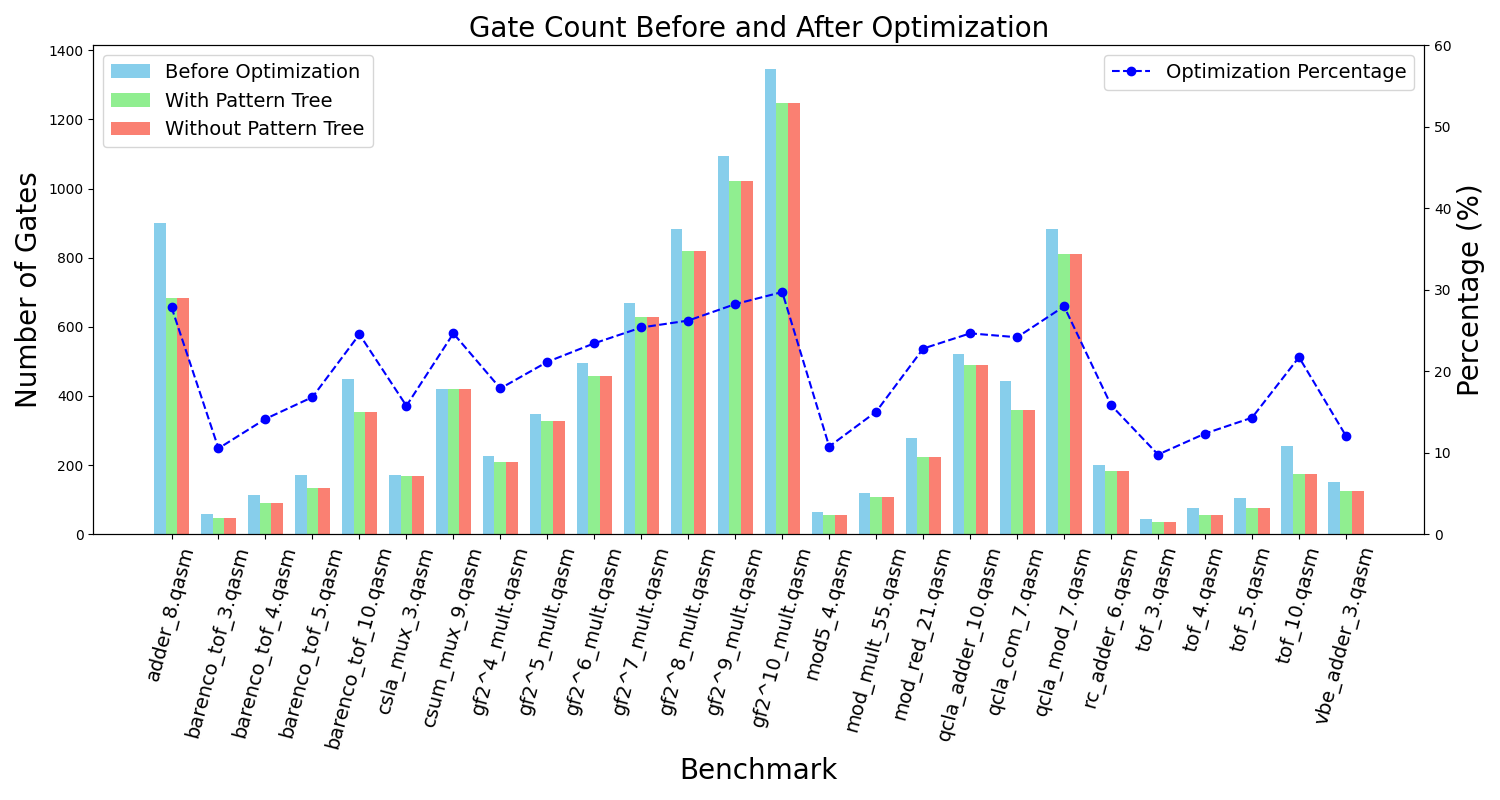}
        \label{fig:gate-opt}
    }
    \caption{The optimization of pattern-tree-based pattern matching algorithm on compilation time when the number of prefix rule is 8}
    \label{fig:opt}
\end{figure}

\paragraph{Circuit verification} 
Our pattern matching routine ensures the equivalence of circuits before and after pattern transformations by the correctness of the transformation rules, which is ensured by the automatic verification of Quartz. However, our preprocessing section currently relies on manually designed routines, although it is inspired by existing rules. Therefore, we use Qiskit's \kn{unitary\_simulator} to verify all benchmarks with less than 16 qubits. The results in Table.\ref{tbl:eff} shows the correctness for all valid benchmark.

\subsection{Compilation Improvement}
Fig.\ref{fig:opt} shows the optimization effect of our pattern-tree-based pattern library on compilation time. 
Since the optimization mechanism of existing approaches differ and no existing work focus on compilation time optimization, we do not make comparison in our work.

Here we set the number of prefix rules to 1\% of all prefix rule candidates, namely 8. 
Fig.\ref{fig:time-opt} shows the compilation time for different benchmarks with and without pattern-tree. For the selected benchmarks, the optimization rate reaches at least 10\%, with an average of 20\%.

To ensure that optimization effect on quantum circuit unchanged after using pattern-tree, we present the optimization results for quantum circuits with and without the pattern tree in Fig.\ref{fig:gate-opt}. The results indicate that using the pattern tree in our framework does not impact its optimization effectiveness.

\subsection{RQ2: The number of prefix rules}

In this subsection, we try to answer RQ2: What is the appropriate number of prefix rules to add to the pattern library?

\begin{figure}[t]
    \subfigure[adder\_8.qasm]{
        \centering
        \includegraphics[width=0.44\textwidth]{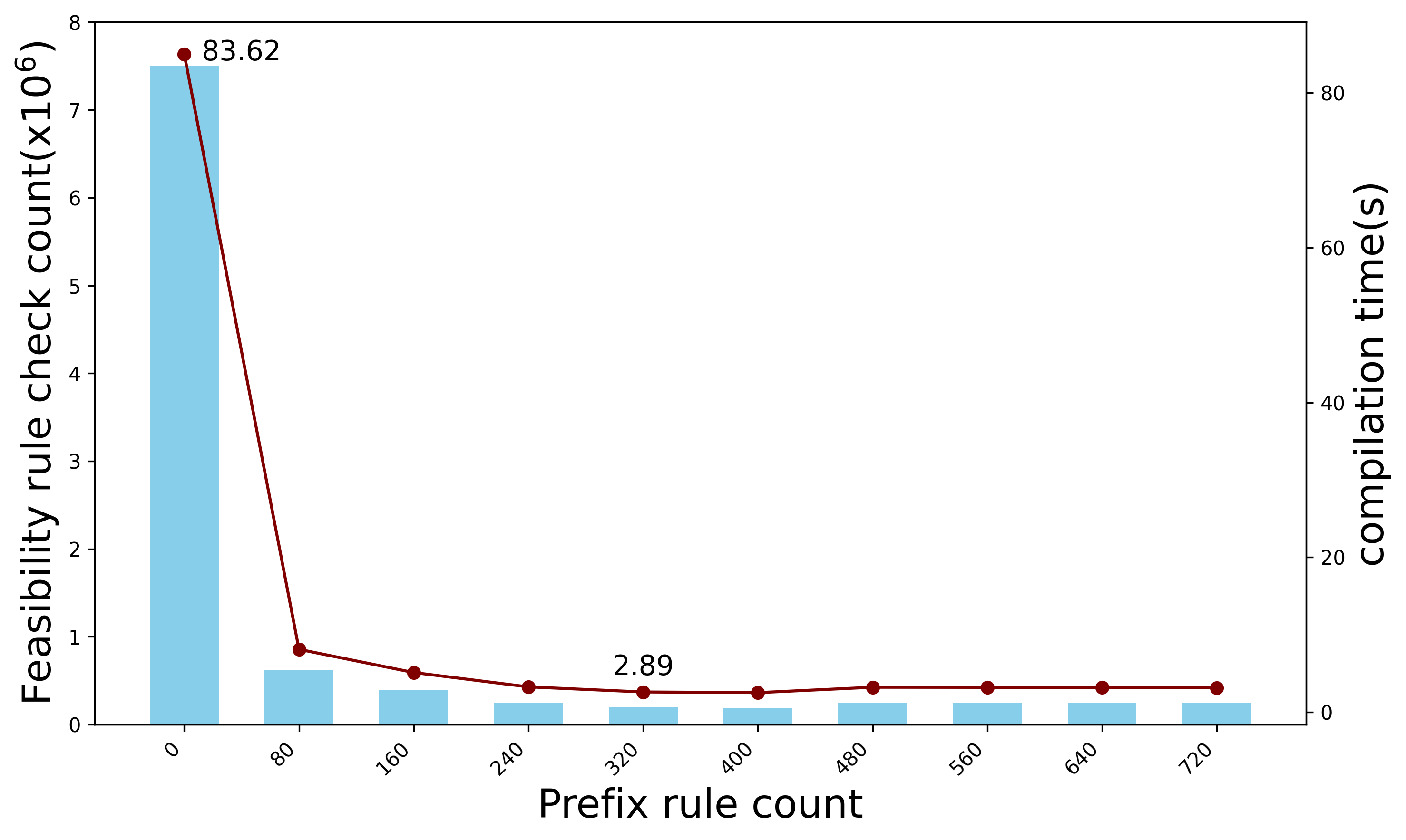}
        \label{fig:adder}
    }
    \subfigure[qcla\_mod\_7.qasm]{
        \centering
        \includegraphics[width=0.44\textwidth]{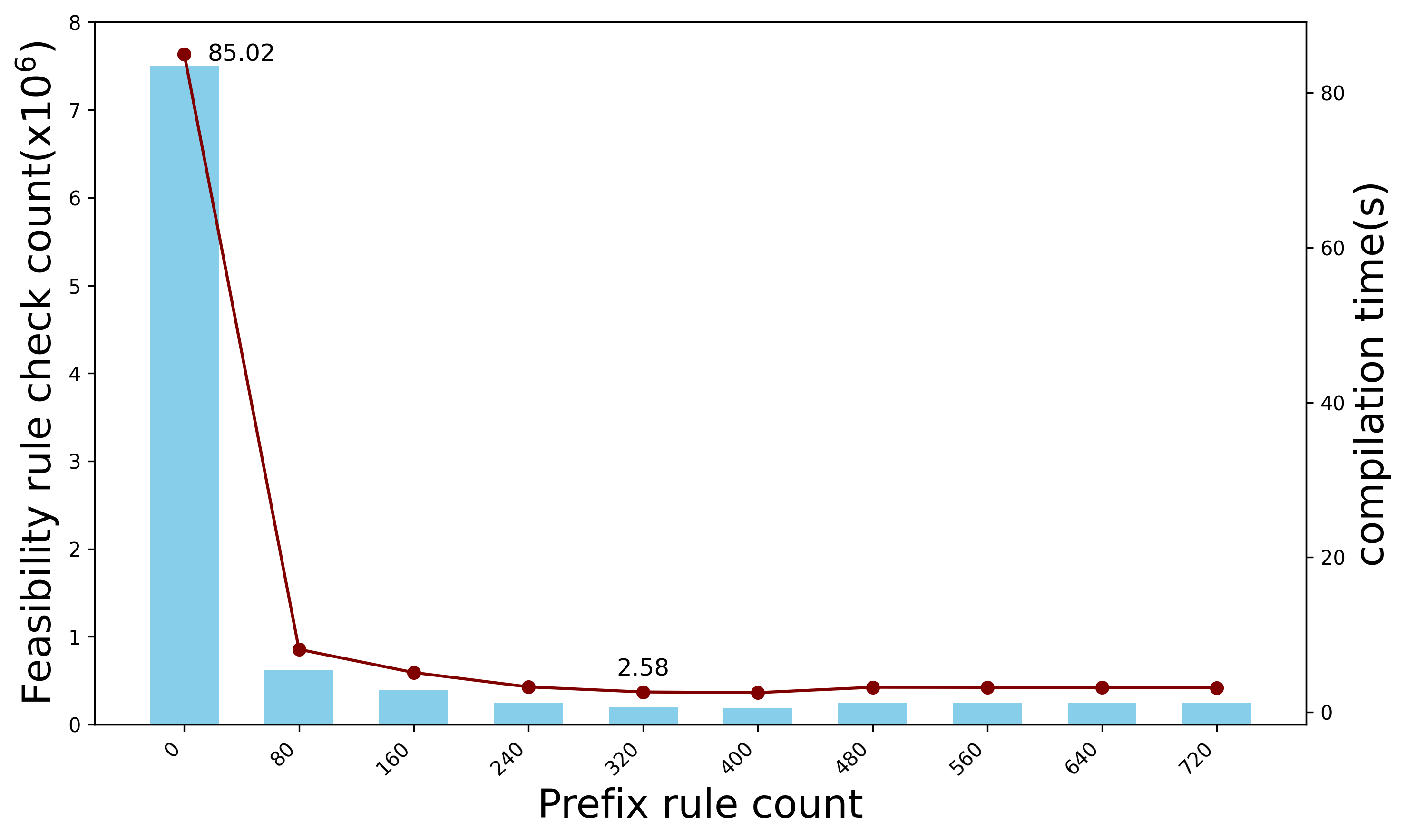}
        \label{fig:qcla}
    }
    \caption{The compilation time and the feasibility rule check count with pattern tree. The left y-axis represents the number of feasibility rule checks, while the right y-axis represents the optimization time of the quantum program.}
    \label{fig:data}
\end{figure}

Since QCiR\cite{chen2022qcir} employs a greedy mechanism to apply transformation rules, it requires all candidate pairs composed of the gates in source of the transformation rule and the gates in input circuit. In QCiR's matching routine, several feasibility rules are used to check the legality of candidate pair. As a result, QCiR’s optimization for each quantum program involves a large number of feasibility rule checks. From Fig. \ref{fig:adder} and Fig.\ref{fig:qcla}, it can be observed that, without using prefix rules, for programs \kn{adder\_8.qasm} and \kn{qcla\_mod\_7.qasm} with approximately 900 quantum gates and nearly 2,000 transformation rules, the number of feasibility rule checks is around 7x10$^6$.

As mentioned in Sec.\ref{sec:motv}, using prefix rules can reduce redundant matches during the execution of the pattern-matching algorithm. In our implementation, redundant matches refer to redundant feasibility rule checks. Therefore, as we can observe from the Fig.\ref{fig:adder} and Fig.\ref{fig:qcla}, after applying the prefix rules, both the compilation time and the number of feasibility rule checks are reduced.
Since most of the generated rules are not used, the pattern-tree method can preemptively exclude these rules, thereby improving the compilation process.
We can observe that when the number of prefix rules reaches around 40\% of the total candidates, the change in compilation time starts to level off. At this point, the optimization of both the number of feasibility rule checks and the compilation time exceeds 90\%.


\section{Related Work}
\label{sec:rel}
\paragraph{Quantum circuit optimization} 
In recent years, quantum circuit optimization becomes one of the most promising research directions in quantum computing. 
Various optimization techniques are developed, including assertion-based optimization\cite{haner2020assertion}, pulse-aware instruction aggregation\cite{shi2019optimized}.
The pattern-matching approach stands out due to its generality and its independence from both the frontend and backend.
Nam \etal\cite{nam2018automated} provides a heuristic quantum circuit optimization algorithm for $\{$\kn{H}, \kn{X}, \kn{CX}, \Rz$\}$.
Quartz\cite{xu2022quartz} automatically generates and verify a comprehensive transformation rule set for a specific gate set, and apply them with a searching algorithm with hyper-parameter.
QUESO\cite{xu2023synthesizing} defines symbolic quantum circuit, and synthesize non-local transformation rules by a path-sum-based method. It applies these rules by using beam search.
The latest Quarl\cite{li2024quarl} utilize a learning-based optimizer to train a gate-selecting policy and a transformation-selecting policy, so that addresses the challenges of the application of cost-preserving rules and cost-increasing rules.
Although our implementation is rule-based since it doesn't explore the whole search space including all equivalent quantum circuits, it can be extended to search-based algorithm.
In our opinion, rule-based algorithm, which has lower computation-cost than search-based algorithm, is not fully developed.

Most of the aforementioned works do not prioritize compilation time. However, variational algorithms\cite{tilly2021variational} are sensitive to compilation time, as they rely on multiple iterations to determine parameters. Gokhale \etal\cite{gokhale2019partial} explore reducing quantum circuit compilation time by reusing portions of previous compilations. 
In contrast, our approach provides a more general optimization for PMT-based quantum compilation time.

\paragraph{Pattern matching}
Pattern matching techniques have been developed in classical compiler optimizations. MLIR\cite{lattner2021mlir}, as a project under the LLVM project\cite{lattner2004llvm}, offers an independent and generic pattern matching and rewriting tool based on a DAG rewriter mechanism. MLIR is also employed in many quantum compilation framework like Quingo\cite{team2020quingo} and QIRO\cite{ittah2022qiro}, although they do not intend to use the pattern matching tool.
Our work is, to some extent, inspired by Espindola \etal's work\cite{espindola2023source}, which proposes a source code and MLIR-based idiom matching and rewriting mechanism.
\section{Conclusion}
In this work, we have presented a methodology called pattern tree based on the observations about relations between transformation rules in quantum circuit.
It aggregates multiple transformation rules which have relation with each other to eliminate the redundancy in the matching procedure and speeding up pattern matching algorithm.
Our experiments demonstrate that pattern-tree works for a widely accpeted benchmark set. Our method can provide at least 10\% optimization on compilation time, and by further exploring building pattern tree, the optimization rate can be increased to 90\%.


\bibliographystyle{plain}
\bibliography{quantum}

\end{document}